\documentclass{elsart}
\def\sun{\odot}
\def\aap{Astron. Astrophys.\,  }
\def\aj{Astronom. J.  }
\def\apj{Astrophys. J. \,  }
\def\apjl{Astrophys. J. L.   \,  }

\def\mnras{Mon. Notices R. astr. Soc. \,  }

\usepackage{natbib}

\usepackage{epsfig}

\usepackage{amssymb}

\begin{document}

\begin{frontmatter}



\title
{
On the three-dimensional   structure of the
nebula around Eta Carinae
}
\author
{L. Zaninetti             }
\address
{
Dipartimento di Fisica Generale, \\
           Via Pietro Giuria 1,   \\
           10125 Torino, Italy
}
\ead{zaninetti@ph.unito.it}

\begin{abstract}

The asymmetric shape  of the nebula around
 $\eta$-Carinae (Homunculus)
  can be explained  by  a spherical expansion in a non-homogeneous medium. Two models
are analyzed: an exponential and an inverse power law dependence
for the density as a function of distance from the equatorial
plane. The presence of a medium with variable density along the
polar direction progressively converts the original spherical
shell into a bipolar nebula. In the case of the nebula around
 $\eta$-Carinae, we know the time elapsed since the great outburst in
1840. An exact match between observed radii and velocities can be
obtained by fine tuning the parameters involved, such as initial
radius, initial velocity and the typical scale that characterizes
the gradient in density. The observed radius and velocity of the
Homunculus as a function of the polar angle in spherical
coordinates can be compared with the corresponding simulated data
by introducing the efficiency in a single or multiple directions.
Once the 3D spatial structure of the Homunculus is obtained, we
can compose the
image by integrating
 along the line
of sight. In order to simulate the observed image, we have
considered a bipolar nebula with constant thickness and
an optically thin emitting layer. Some simulated cuts of the
relative intensity are reported and may represent a useful
reference  for the astronomical cuts.
\end{abstract}
\end{frontmatter}

\parindent=0.5 cm

\section{Introduction}

The nebula around $\eta$-Carinae  was
discovered by \cite{Thackeray1949}
and  the
name ``the Homunculus'' arises from the fact
that  on the photographic plates it
resembled a small plump man, see \cite{Gaviola1950}.
More details on the various aspects of
 $\eta$-Carinae can be found in \cite{Smith2009}.
The structure  of the Homunculus Nebula
around $\eta$-Carinae has  been  analyzed
with different models,
we cite  some of them:
\begin{itemize}
\item The shape  and kinematics is explained by the interaction
of the winds expelled by the central star at different injection
velocities, see \cite{Icke1988}. \item The  possibility that the
nebulae around luminous blue variables (LBVs) are shaped by
interacting winds  has  been analyzed by \cite{Nota1995}. In this
case  a  density contrast profile of the form $\rho$ =
$\rho_0 (1 + 5  \cos^4 \Theta ) $
was used
where $\Theta$ is the angle to
the equatorial plane. \item The origin and evolution of the
bipolar nebula has  been modeled by a  numerical two-dimensional
gasdynamic model
 where  a stellar wind interacts with an
aspherical circumstellar environment, see \cite{Frank1995}. \item
Cooling models form  ballistic
 flows (that is, a
pair of cones each with a spherical base) whose lateral edges
become wrinkled by shear instabilities, see \cite{Dwarkadas1998}.
\item The scaling relations derived from the theory of radiatively
driven winds can model the outflows from luminous blue variable
(LBV) stars, taking account of stellar rotation and the associated
latitudinal variation of the stellar flux due to gravity
darkening. In particular for  a star rotating close to its
critical speed, the decrease in effective gravity near the equator
and the associated decrease in the equatorial wind
 speed results
naturally in a bipolar, prolate interaction front, and therefore
in an asymmetric wind, see  \cite{Dwarkadas2002}.
 \item Two
oppositely ejected jets inflate two lobes (or bubbles)
representing a unified model for the formation of bipolar lobes,
see \cite{Soker2004,Soker2007}. \item A two-dimensional,
time-dependent hydrodynamical simulation of radiative cooling, see
\cite{Gonzales_2004}. \item Launch of  material normal to the
surface of the oblate rotating star with an initial kick velocity
that scales approximately with the local escape speed, see
\cite{Smith_2007}. \item
 A  3D model of wind-wind collision
for  X-ray emission from a supermassive star,
see \cite{Parkin2009}.
\item
Two-dimensional hydrodynamical simulations of
the eruptive events of the 1840s (the great outburst)
and 1890s (the minor outburst), see \cite{Gonzales2010}.
\end{itemize}
The models cited leave some questions
unanswered or only partially answered:
\begin {itemize}
\item  Which is  the law of motion which  regulates
       the expansion?
\item  Is it possible to model the complex
       three-dimensional (3D)
       behaviour of
       the velocity field of the expanding nebula?
\item  Is it possible to make an
       evaluation of the reliability of the
       numerical results
       on radius and velocity compared to
       observed values?
\item  Is it possible to evaluate
       the intensity of the $H_2$ image
       of the nebula?
\item  Is it  possible to
       build
       cuts of the model intensity which can be compared
       with existing
       observations?
\end{itemize}

In order  to answer these questions, Section~\ref{sec_eta}
describes three observed morphologies of PNs,
Section~\ref{sec_motion} analyzes two different laws of motion
with their associated velocities
which  model the aspherical expansion and Section~\ref{sec_image}
contains detailed information on how to build the image of the
Homunculus as well as some sectional
cuts through the relative intensity of emission.

\section{Homunculus properties}

\label{sec_eta}

This Section reviews the basic data of the Homunculus
as well as an ad hoc formula for the latitude-dependent
wind.
The star $\eta$-Carinae had a great outburst in 1840
and at the moment of writing presents a bipolar shape
called the Homunculus,
its distance is 2250 pc, see \cite{Smith2002}.
A more refined classification  distinguishes
between the large and little Homunculus, see \cite{Gonzales_2006}.
The Homunculus has been observed at different
 wavelengths such as the
ultraviolet and infrared  by \cite{Smith_2004,Smith2009}, x-ray
by \cite{Corcoran2004}, $[FeII]\lambda$16435 by \cite{Smith_2005},
ammonia by \cite{Smith_2006}, radio-continuum by
\cite{Gonzales_2006}, near-infrared by \cite{Teodoro_2008} and
scandium and   chromium lines  by \cite{Bautista_2009}.

\begin{figure}
  \begin{center}
\includegraphics[width=10cm]{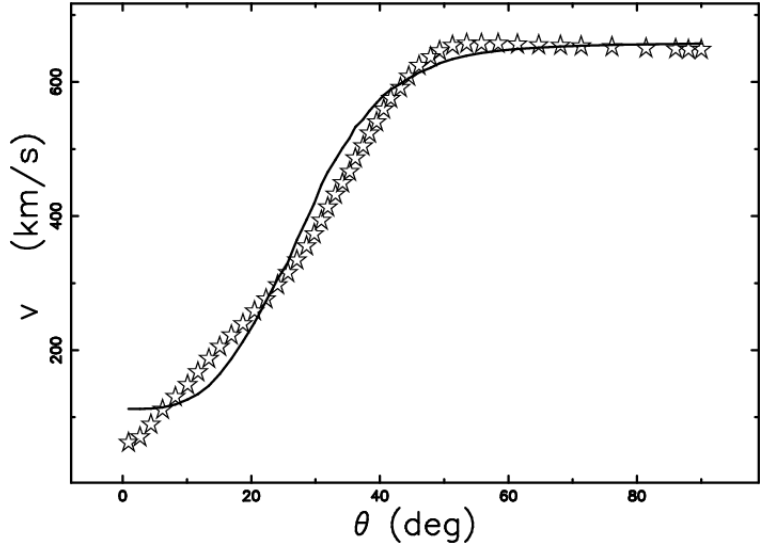}
  \end {center}
\caption {
Expansion velocity  versus latitude of  the
$H_2$ shell  (Smith 2006, open stars)
and fit as given by  formula~(\ref{vgonzales}) (full line).
The fitting parameters are the same as  \cite{Gonzales2010}:
$\lambda$=1.9, $v_1=670\,km/s$ and $v_2=100\, km/s$.
          }%
    \label{gonzales}
    \end{figure}

 Referring to Table 1 in~\cite{Smith2006}, we can
fix the major radius at  22014~AU (0.106~pc)
and the equatorial radius
at 2100~AU (0.01~pc).
The expansion velocity rises from $\approx ~ 93~km/s$
at the equator to $\approx ~ 648~km/s$  in the polar
direction, see Table~1 and Figure~4 in~\cite{Smith2006}.
The thickness of the $H_2$ shell is roughly
$2-3\%$  of the  polar radius, see \cite{Smith2006}.
The expansion  speed of the outer  $H_2$ shell
of $\eta$-Carinae   has  been fitted with the following latitude
dependent velocity as given by
\begin{equation}
v=
\frac{
{\it v_1}\, \left( {\it v_2}+{{\rm e}^{2\,{\it \lambda}\,\cos \left( 2\,
\Theta \right) }}{\it v_1} \right)
}
{
{\it v_1}\, \left( 1+{{\rm e}^{2\,{\it \lambda}\,\cos \left( 2\,\Theta
 \right) }} \right)
}
\quad,
\label{vgonzales}
\end{equation}
where the parameter $\lambda$ controls the shape of the Homunculus,
$\Theta$ is the polar angle;
$v_1$ and $v_2$ are the velocities in the polar
and equatorial direction,
see \cite{Gonzales2010}.
Figure \ref{gonzales} reports the data of \cite{Smith2006}
as well the fit of \cite{Gonzales2010}.

\section{Law of motion}

\label{sec_motion} This Section presents a standard solution for
the law of motion in an interstellar medium with constant density.
The assumption  of constant density can be generalized considering
a medium   which presents  a decreasing behaviour  in density
with
increasing distance from the main plane of symmetry: in particular
we considered an exponential and a power law decrease in density.
As a consequence two new laws of motion are deduced. In all three
cases, the velocity is deduced assuming  the conservation of
momentum   and has a simple analytical
form.
The  resulting  first order equations for the radius
have separable variables and therefore the solution can be found.
In the first case  is possible to derive an analytical expression
for the radius  and in the two other cases   there exists
 an implicit
equation for the radius.
A subsection deals with the simulation quality
as well as along
many directions.

\subsection{Spherical Symmetry - Conservation of Momentum}

The thin layer approximation assumes
that all the swept-up
gas accumulates in an infinitely
thin shell just behind
the shock front.
The conservation of radial momentum
at distance $R$,$P(R)$,
requires that
\begin{equation}
P(R) = P_0(R_0)
\quad ,
\end{equation}
where $R_0$   is   the initial radius and $P_0(R_0)$ the radial
momentum at  $R_0$ evaluated when  $t=t_0$.
The conservation of
the momentum gives
\begin{equation}
\frac{4}{3} \pi R^3 \rho \dot {R} = P_0(R_0)
\quad,
\end{equation}
where  $\dot{R}$ is the  velocity
of the advancing shock,
see \cite{Dyson1997,Padmanabhan_II_2001}
and  $\rho$
is the density of the ambient medium.
The law of motion is
\begin{equation}
R = R_0 \left  ( 1 +4 \frac{\dot {R_0}} {R_0}(t-t_0) \right )^{\frac{1}{4}}
\label{radiusm}
\quad .
\end{equation}
and the velocity
\begin{equation}
\dot {R} = \dot {R_0} \left ( 1 +4 \frac{\dot {R_0}} {R_0}(t-t_0)\right )^{-\frac{3}{4}}
\label{velocitym}
\quad .
\end{equation}
We can derive $\dot {R_0}$ from equation (\ref{radiusm})
and insert it in equation (\ref{velocitym})
\begin{equation}
\dot {R} =\frac{1}{4(t-t_0)}  \frac{R^4-R_0^4}{R_0^3}
\left  ( 1+\frac{R^4-R_0^4}{R_0^4} \right )^{-\frac{3}{4}}
\label{velocitym2}
\quad .
\end{equation}
The astrophysical  units are:  $t_4$  and
$t_{0,4}$
which  are $t$ and  $t_0$
expressed  in $10^4$ \mbox{yr} units,
$R_{pc}$ and $R_{0,pc}$ which are
$R$ and  $R_0$  expressed in  pc,
$\dot {R}_{kms}$
and
$\dot {R}_{0,kms}$
which are
 $\dot{R} $ and  $\dot{R}_0$
expressed
in $km/s$.
Therefore the previous formula becomes
\begin{equation}
\dot {R}_{kms} =24.49 \frac{1}{(t_4-t_{0,4} )}
\frac{R_{pc}^4-R_{0,pc}^4}
{R_{0,pc} ^3}
\left  ( 1+\frac{R_{pc}^4-R_{0,pc}^4}{R_{0,pc}^4} \right )^{-\frac{3}{4}}
\label{velocitym2astro}
\quad .
\end{equation}
An interesting quantity   can be  the swept mass
at  a distance  $R$ from the origin
\begin{equation}
M = \frac {4} {3} \pi R^3 \rho
\quad  .
\end{equation}
The  density can be
\begin{equation}
\rho = 1.4 n_1  m_H  \frac {g} {cm^3}  = 1.4 \times 1.66 10^{-24}
 n_1 \frac {g} {cm^3}
\quad ,
\end{equation}
where $n_1$
is   the
number density  expressed  in units of   particle
$\mathrm{cm}^{-3}$
and the factor 1.4 takes into account
elements heavier than hydrogen.
Introducing the solar mass, $M_{\sun}$,
the    swept-up
  mass is
\begin{equation}
M =
0.143 \,{{\it R_1}}^{3}{\it n_1}  M_{\sun}
\quad  ,
\end{equation}
where $R_1$ is the radius  expressed in  pc
units.

\subsection{Exponentially varying medium }
The number density of an exponentially varying medium is described by
\begin{equation}
n (z) = n_0 \exp {- \frac {z}{h} }
\quad  ,
\label{exponential}
\end{equation}
where  $z$ is the distance from the equatorial plane,
$n_0$ is the number
of particles at $R=R_0$ and $h$ is the scale height.

\begin{figure}
  \begin{center}
\includegraphics[width=10cm]{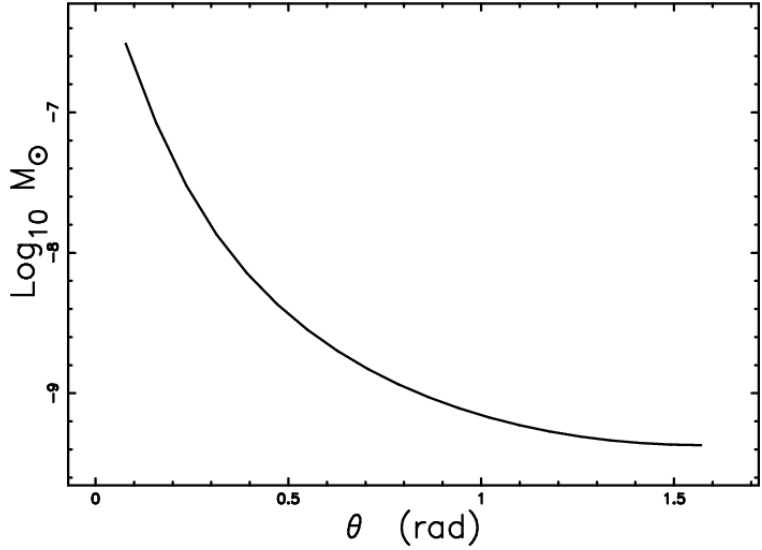}
  \end {center}
\caption {
Decimal logarithm  of swept-up mass when
$R=0.05\,pc$, $n_1=1$
 and
$\Delta\;\Omega=1$.
Physical parameters as in Table~\ref{parametershom}.
          }%
    \label{mass}
    \end{figure}

A three-dimensional (3D) expansion will
be characterized by the following
properties
\begin {itemize}
\item Dependence from  the instantaneous radius of the shell
      on the latitude  angle $\theta$ which has a range
      $[-90 ^{\circ}  \leftrightarrow  +90 ^{\circ} ]$.

\item Independence of the instantaneous radius of the shell
      from  $\phi$, the azimuthal  angle  in the x-y  plane,
      which has a range
      $[0 ^{\circ}  \leftrightarrow  360 ^{\circ} ]$.
\end {itemize}
The mass, $M$,   swept-up along a  solid angle
$ \Delta\;\Omega $,  between 0 and $R$ is
\begin{equation}
M(R)=
\frac { \Delta\;\Omega } {3} 1.4 \, m_H n_0 I_m(R)
+ \frac{4}{3} \pi R_0^3 n_0 m_H \,1.4
\quad  ,
\end {equation}
where
$m_H$ is the mass of hydrogen and
\begin{equation}
I_m(R)  = \int_{R_0} ^R r^2 \exp ({ - \frac {r \sin (\theta) }{ h}  }) dr
\quad ,
\end{equation}
where $R_0$ is the initial radius.
The integral is
\begin{eqnarray}
I_m(R)  =
\frac
{
h \left( 2\,{h}^{2}+2\,R_0h\sin \left( \theta \right) +{R_0}^{2} \left(
\sin \left( \theta \right)  \right) ^{2} \right) {{\rm e}^{-{\frac {R_0
\sin \left( \theta \right) }{h}}}}
}
{
\left( \sin \left( \theta \right)  \right) ^{3}
}
           \nonumber\\
- \frac
{
h \left( 2\,{h}^{2}+2\,Rh\sin \left( \theta \right) +{R}^{2} \left(
\sin \left( \theta \right)  \right) ^{2} \right) {{\rm e}^{-{\frac {R
\sin \left( \theta \right) }{h}}}}
}
{
\left( \sin \left( \theta \right)  \right) ^{3}
}
\quad .
\end{eqnarray}
In this case, the swept mass is a function of the latitude
angle $\theta$ and can be plotted assuming
$\Delta\;\Omega=1$, see  Figure \ref{mass}.

This  numerical evaluation gives a  simple explanation for the
asymmetry of the nebula around  $\eta$-Carinae: a smaller mass being swept
means a greater velocity of the advancing radius of the nebula.

Conservation of momentum gives
\begin{equation}
M(R)   \dot {R}=
M(R_0) \dot {R_0}
\quad  .
\label{eqnmomentum}
\end{equation}
In this differential equation of the first order in $R$, the
variables can be separated and an integration
term-by-term gives
\begin{equation}
\int_{R_0}^{R}  M(r)  dr =
M(R_0) \dot {R_0} \times ( t-t_0)
\quad  ,
\label{eqnintegration}
\end{equation}
where  $t$ is the time and $t_0$ is the time at $R_0$.
The resulting non-linear equation ${\mathcal{F}}_{NL}$,
is equation (17) in \cite{Zaninetti2009a}.
The radius
$R_{pc}$
as function of the time,  is found by a numerical method, in the case of
 $\eta$-Carinae  the age is $158~yr$  and therefore
$t_4-t_{0,4}=t_0/(10^4 yr)$=0.0158.

\subsection{Power law  medium}

A  possible form for a power law profile of the medium surrounding
the Homunculus nebula is
\begin{equation}
n (z) = n_0  \bigl ( \frac{z}{R_0} \bigr )^{-\alpha}
\quad  ,
\label{powerlaw}
\end{equation}
where  $z=R\times \sin ( \theta) $ is the distance from the equatorial plane,
$R$ is the instantaneous radius of expansion,
$n_0$ is the number
of particles at $R=R_0$
and $\alpha$ is a coefficient $>0$.

The  swept-up   mass, $M$,  along  a solid angle $ \Delta\;\Omega $
between 0 and $R$ is
\begin{equation}
M(R)=
\frac { \Delta\;\Omega } {3} 1.4 \,  m_H n_0 I_m(R)
+ \frac{4}{3} \pi R_0^3 n_0 m_H \,1.4
\quad  ,
\end {equation}
where
\begin{equation}
I_m(R)  = \int_{R_0} ^R r^2
\bigl ( \frac {r \sin (\theta) }{ R_0} \bigr )^{-\alpha}   dr
\quad ,
\end{equation}
where $R_0$ is the initial radius.
 Integrating gives:
\begin{eqnarray}
I_m(R)  =
\frac
{
{R}^{3} \left( {\frac {R\sin \left( \theta \right) }{{\it R_0}}}
 \right) ^{-\alpha}
} { 3-\alpha } \quad .
\end{eqnarray}
The resulting non-linear equation ${\mathcal{F}}_{NL}$
of motion
expressed in astrophysical units
can obtained by eqns. (\ref{eqnmomentum}) and  (\ref{eqnintegration})
and is
\begin{eqnarray}
{\mathcal{F}}_{NL} =
- \,{{\it R_{0,pc}}}^{4}{\alpha}^{2}+{\it R_{pc}}\,{{\it R_{0,pc}}}^{3}{\alpha
}^{2}+{\it R_{pc}}\,{{\it R_{0,pc}}}^{3} \left( \sin \left( \theta \right)
 \right) ^{- \,\alpha}\alpha
\nonumber \\
+ 7.0\,{{\it R_{0,pc}}}^{4}\alpha- \,{{
\it R_{0,pc}}}^{4} \left( \sin \left( \theta \right)  \right) ^{- \,
\alpha}\alpha- 7.0\,{\it R_{pc}}\,{{\it R_{0,pc}}}^{3}\alpha
\nonumber  \\
+ 3\,{{\it R_{0,pc}
}}^{4} \left( \sin \left( \theta \right)  \right) ^{- \,\alpha}-
 12\,{{\it R_{0,pc}}}^{4}- 4.0\,{\it R_{pc}}\,{{\it R_{0,pc}}}^{3} \left( \sin
 \left( \theta \right)  \right) ^{- \,\alpha}
\nonumber \\
+ \,{{\it R_{pc}}}^{-
 \,\alpha+ 4.0}{{\it R_{0,pc}}}^{\alpha} \left( \sin \left( \theta
 \right)  \right) ^{- \,\alpha}+ 12\,{\it R_{pc}}\,{{\it R_{0,pc}}}^{3}
\nonumber \\ -
 0.122\,{{\it R_{0,pc}}}^{3}{\it { \dot{R}_{0,kms}} }\, \left
( t_4-{\it t_{0,4}}
 \right)
\nonumber \\
- 0.01\,{{\it R_{0,pc}}}^{3}{\it { \dot{R}_{0,kms}} }\,
\left( t_4-{
\it t_{0,4}} \right) {\alpha}^{2}
\nonumber  \\
+ 0.0714\,{{\it R_{0,pc}}}^{3}{\it
{ \dot{R}_{0,kms}} }\, \left( t_4-{\it t_{0,4}} \right) \alpha
=0
\quad ,
\label{nl_power}
\end{eqnarray}
where $t_4$
and
$t_{0,4}$
 are $t$ and  $t_0$
expressed in $10^4$ \mbox{yr} units,
$R_{pc}$ and $R_{0,pc}$  are
$R$ and  $R_0$  expressed in  pc,
$\dot {R}_{kms}$
and
$\dot {R}_{0,kms}$
are
$\dot{R} $ and  $\dot{R}_0$   expressed
in $km/s$ and
$\theta$ is expressed in radians.

It is not possible to find $R_{pc}$ analytically and
a numerical method must be implemented.
In  our case, in order
to find  the root of  ${\mathcal{F}}_{NL}$,
the FORTRAN SUBROUTINE ZRIDDR from \cite{press} has been used.
The unknown parameters,
$R_{0,pc}$  and
$\dot {R}_{0,kms}$, are found from different runs
of the code, $t_4-t_{0,4}$ is an input parameter.

\subsection{ Quality of the simulations}

 As in \cite{Zaninetti2009a}, two parameters
are introduced to assess the quality of the simulations.
The first, $\epsilon$,
compares observed and simulated quantities:
\begin{equation}
\epsilon  =(1- \frac{\vert( R_{\mathrm {pc,obs}}- R_{pc,\mathrm {num}}) \vert}
{R_{pc,\mathrm {obs}}}) \cdot 100
\,,
\label{efficiency}
\end{equation}
where $R_{pc,\mathrm {obs}}$ is
observed radius, in parsec
and $R_{pc,\mathrm {num}}$ is the radius  from our simulation
in parsec.

The second defines an observational
reliability,
$\epsilon_{\mathrm {obs}}$, over the whole range
of the latitude  angle  $\theta$,
\begin{equation}
\epsilon_{\mathrm {obs}}  =100(1-\frac{\sum_j |R_{pc,\mathrm {obs}}-R_{pc,\mathrm {num}}|_j}{\sum_j
{R_{pc,\mathrm {obs}}}_{,j}})
,
\label{efficiencymany}
\end{equation}
where
the  index $j$  varies  from 1 to the number of
available observations.
Those of the Homunculus are represented by Table 1 in
\cite{Smith2006}.

\subsection{Simulations quality for exponentially varying medium}
A typical set of parameters  which  allows
the Homunculus nebula around $\eta$-Carinae to be simulated
in the presence of a medium whose density
decreases exponentially
is reported in
Table \ref{parametershom}.
Table \ref{efficiency_exp}
presents
numbers concerning the quality of fit.

\begin{figure}
  \begin{center}
\includegraphics[width=10cm]{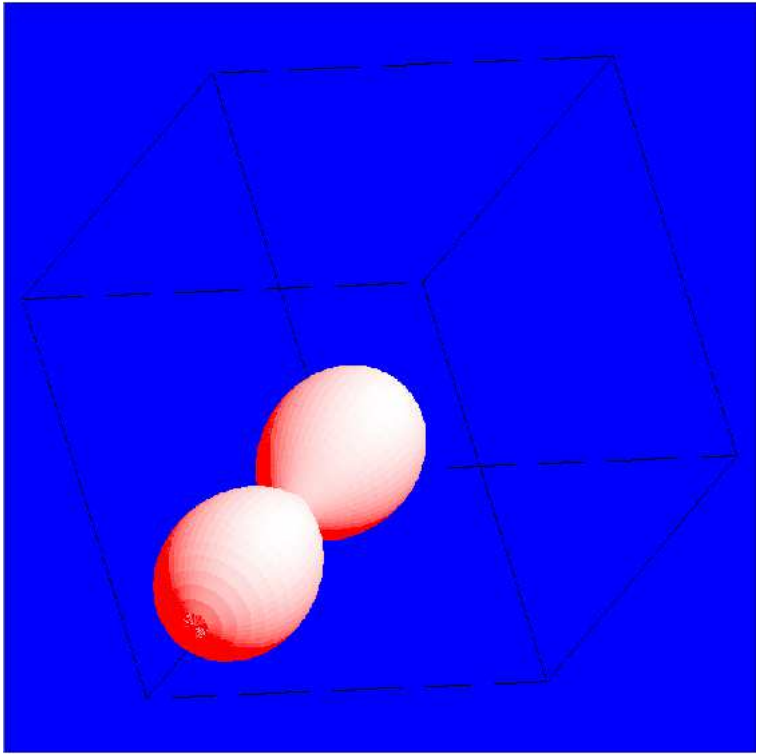}
  \end {center}
\caption {
Simulations lead to this picture
of the Homunculus for an exponentially varying medium.
The orientation  of the figure is characterized by the Euler angles ,
which are
     $ \Phi   $=130$^{\circ }$,
     $ \Theta $=40$^{\circ }$
and  $ \Psi   $=-140$^{\circ }$.
Physical parameters as in Table~\ref{parametershom}.
          }%
    \label{eta_faces}
    \end{figure}

\begin{figure}
  \begin{center}
\includegraphics[width=10cm]{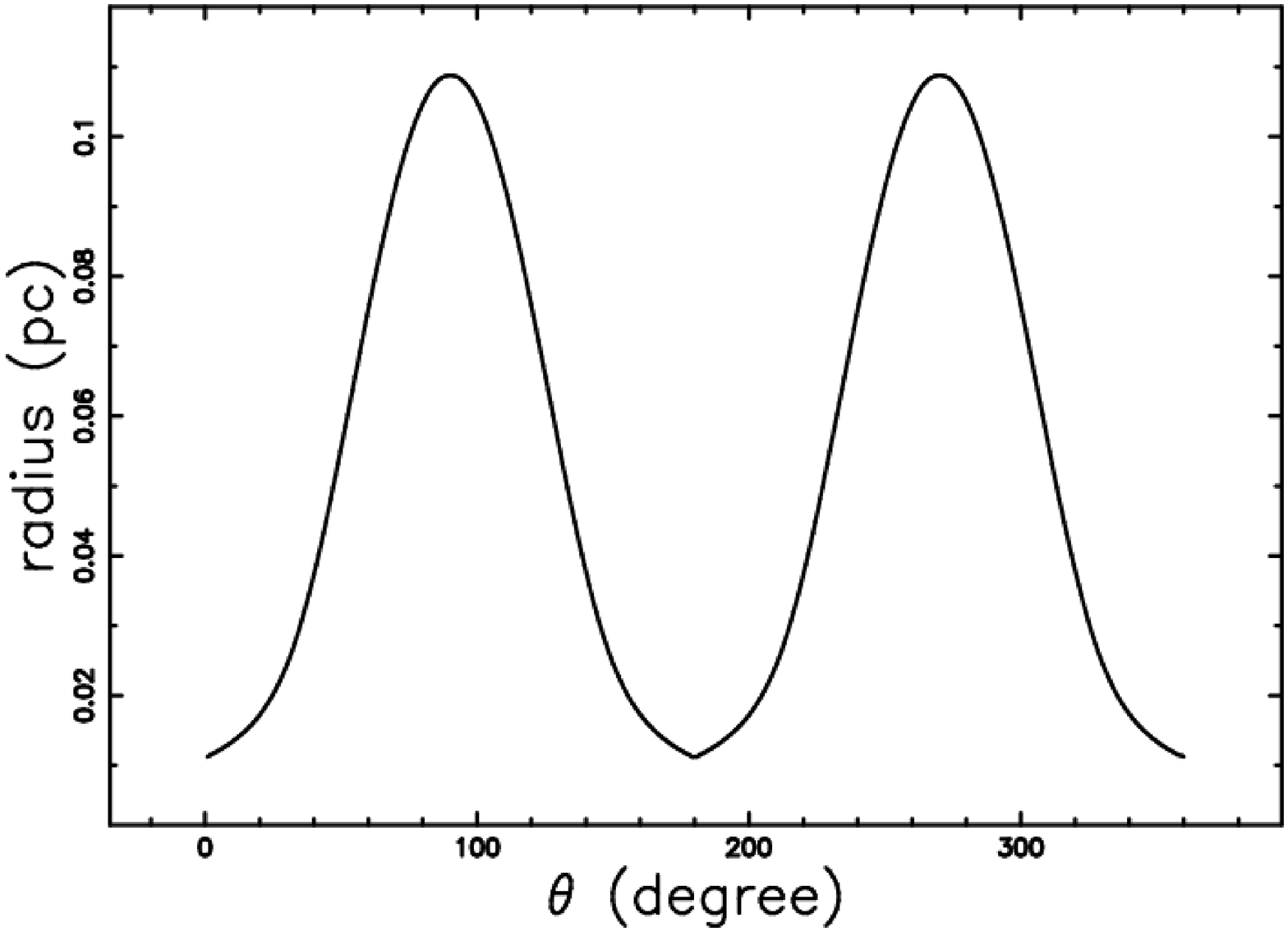}
  \end {center}
\caption {
 Radius as a function of latitude for an
 exponentially varying medium (dotted line)
and  astronomical data with error bar.
Physical parameters as in Table~\ref{parametershom}.
          }%
    \label{eta_radius}
    \end{figure}

\begin{figure}
  \begin{center}
\includegraphics[width=10cm]{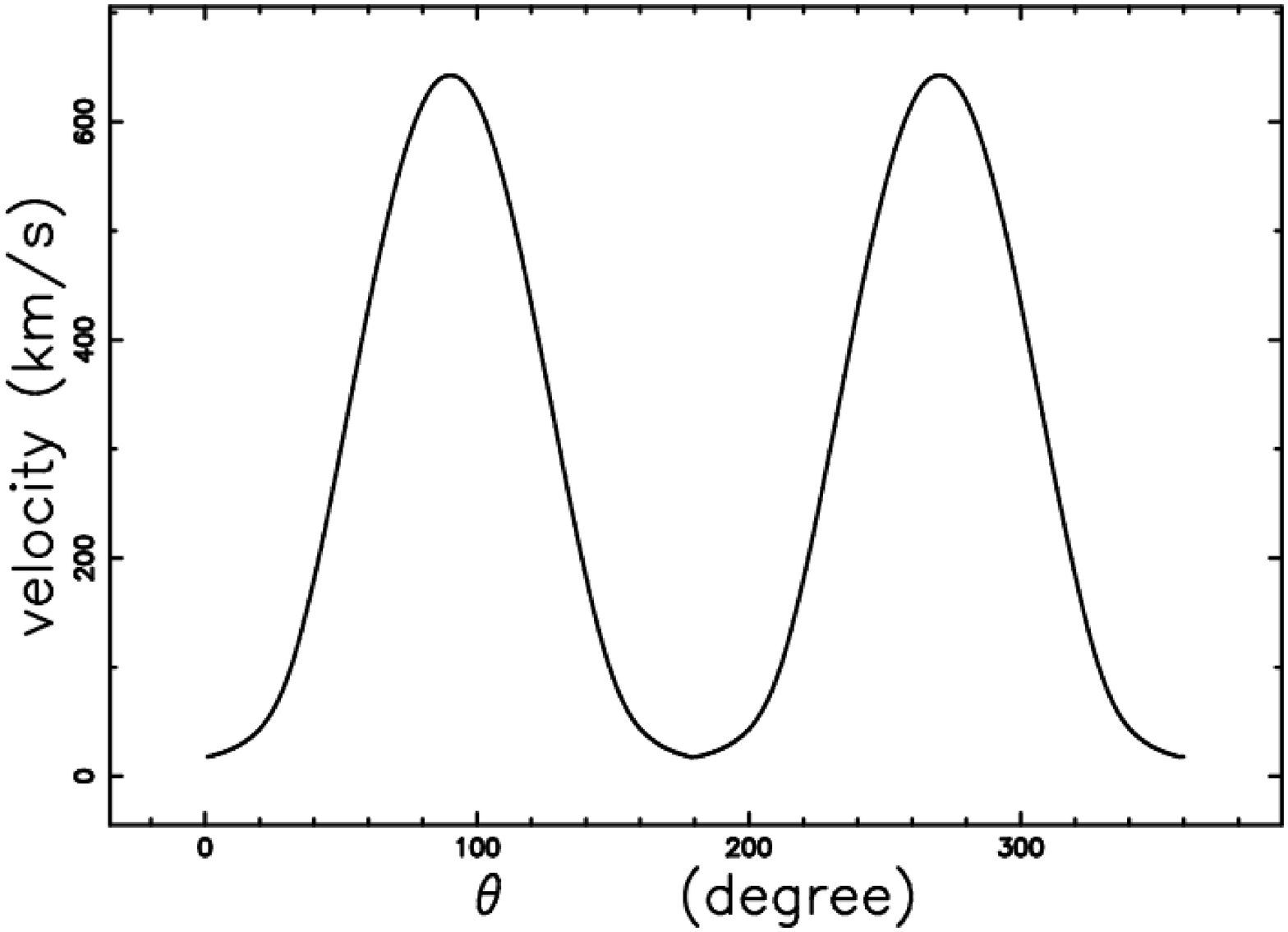}
  \end {center}
\caption {
Velocity as a function of latitude for an
exponentially
 varying medium
(dotted line)
and  astronomical data with error bar
Physical parameters as in Table~\ref{parametershom}.
          }%
    \label{eta_velocity}
    \end{figure}

\begin{table}
      \caption{
 Parameter values used to simulate
the observations of the Homunculus nebula for a medium
varying exponentially (first 4 values) or a power law
(2nd set of 4 values)}
         \label{parametershom}
      \[
         \begin{array}{cc}
            \hline
            \noalign{\smallskip}
 \mbox {Initial ~expansion~velocity, ${\dot R}_{{0,1}}$
 [km~s$^{-1}$}]    & 8 000              \\
 \mbox {Age~($t_4-t_{0,4}$) [10$^4$~yr]} & 0.0158  \\
 \mbox {Scaling~h     [pc] }             & 0.0018 \\
 \mbox {Initial~radius~ $R_0$ ~[pc] }    & 0.001 \\
            \noalign{\smallskip}
            \hline
            \noalign{\smallskip}
 \mbox {Initial ~expansion~velocity, ${\dot R}_{{0,1}}$
 [km~s$^{-1}$}]    & 40,000              \\
\mbox {Age~($t_4-t_{0,4}$) [10$^4$~yr]} & 0.0158 \\
\mbox {Initial~radius~ $R_0$ ~[pc] }    & 0.0002 \\
\mbox {Power~law~coefficient $\alpha$ } & 2.4    \\
            \noalign{\smallskip}
            \hline
            \noalign{\smallskip}
         \end{array}
      \]
   \end{table}
\begin{table}
      \caption{
Agreement between observations and simulations for the Homunculus nebula,
for an exponentially varying medium.
}
         \label{efficiency_exp}
      \[
         \begin{array}{lcc}
            \hline
            \noalign{\smallskip}
~~~                                     & radius   &  velocity    \\
\mbox {$\epsilon$} (\%)-polar~ direction & 97       &  99          \\
\mbox {$\epsilon$} (\%)-equatorial~ direction & 87       &  19          \\
\mbox {$\epsilon_{obs}$} (\%)      &  85   &  75           \\
            \noalign{\smallskip}
            \hline
         \end{array}
      \]
   \end{table}
The bipolar character of the Homunculus is
 shown
in Figure~\ref{eta_faces}.
In order to better visualize the two lobes,
Figure~\ref{eta_radius} and  Figure~\ref{eta_velocity}
show the radius and velocity
as a function of the angular position $\theta$.
The  orientation of the observer is
characterized by the three
Euler   angles
$(\Phi, \Theta, \Psi)$,
see \cite{Goldstein2002};  different  Euler angles
produce different observed shapes.

\begin{figure}
  \begin{center}
\includegraphics[width=5cm]{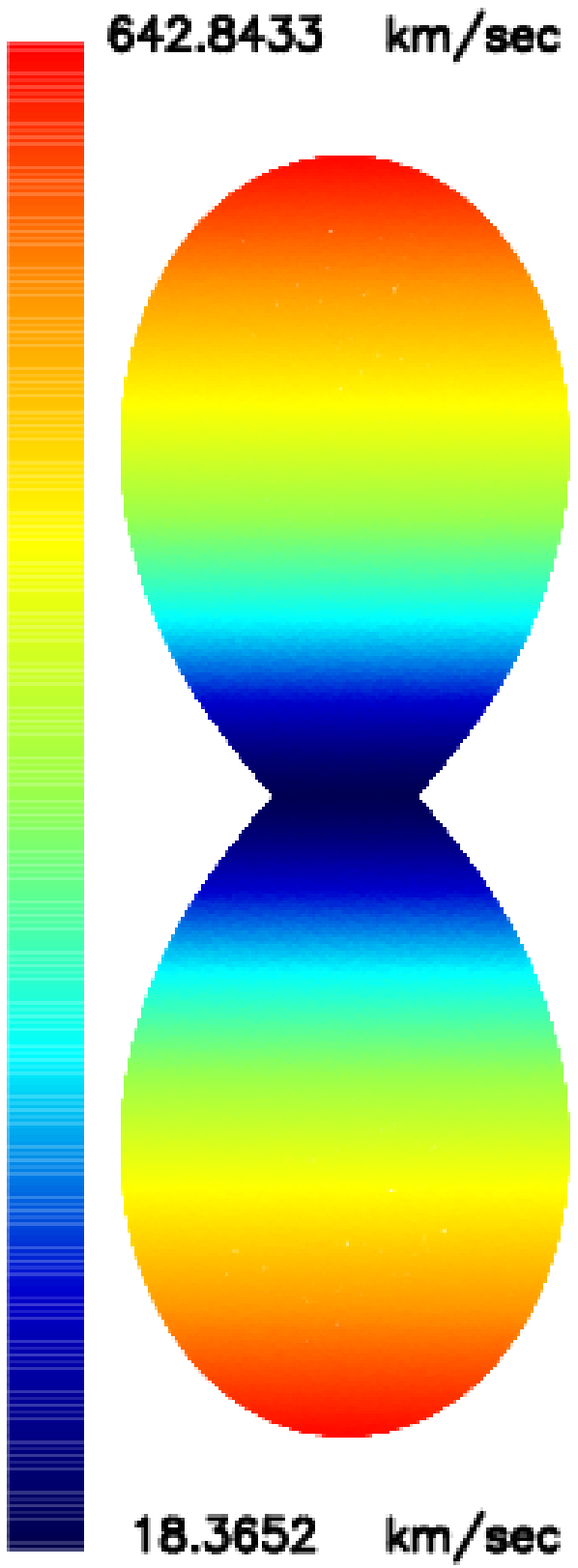}
  \end {center}
\caption {
Map of the expansion velocity for an exponentially
varying medium.
  Physical parameters as in Table~\ref{parametershom}.
          }%
    \label{eta_velocity_field}
    \end{figure}

The velocity field
is  shown in Figure \ref{eta_velocity_field}.

The accuracy with which our code reproduces
the spatial shape and the
velocity field over 18 directions
of the Homunculus nebula as given by
formula~(\ref{efficiencymany}) is reported in Table~\ref{efficiency_exp}.
From a careful analysis of Table~\ref{efficiency_exp} we can conclude that
the spatial shape over 18 directions is well modeled by an exponential medium
, $\epsilon_{obs} = 85 \%$.
The overall efficiency of the field is smaller
$\epsilon_{obs} = 75 \%$.
We can therefore conclude
that formula (\ref{efficiencymany}) which gives
the efficiency  over all the range of polar  angles
represents a better  way  to describe  the results
in  respect to the efficiency  in a single direction
as given by formula (\ref{efficiency}).

\subsection{Simulations quality for power law medium}

For assumed parameters see Table
\ref{parametershom},
Table~\ref{efficiency_power} reports
the accuracy of radius and velocity in two directions.

\begin{table}
      \caption{
 Agreement between model for a power
law medium and observations.
}
         \label{efficiency_power}
      \[
         \begin{array}{lcc}
            \hline
            \noalign{\smallskip}
~~~                                     & radius   &  velocity    \\
\mbox {$\epsilon$} (\%)-polar~ direction & 78       &  76          \\
\mbox {$\epsilon$} (\%)-equatorial~ direction & 6   &  2          \\
\mbox {$\epsilon_{obs}$} (\%)                 & 79   &  73          \\
            \noalign{\smallskip}
            \hline
         \end{array}
      \]
   \end{table}

\section{Image}

\label{sec_image}
An image of an astrophysical object
is composed in a simulation
by combining the intensities that characterize
different points.
For an optically thin medium the transfer equation
provides the emissivity to be multiplied with the distance
in the line of sight.
The transfer
equation has been  analyzed in Section~5.1
of~\cite{Zaninetti2009a}. The Homunculus nebula was observed
through emission-line spectra such as $H_2$ and $[Fe_{II}]$, see
\cite{Smith2006}. We now outline a possible source of radiation.
The volume emission coefficient of the transition $j_{21}$ is
\begin{equation}
j_{21} = \frac{n_2 A_{21} h \nu_{21} } { 4 \pi}
\quad  ,
\end{equation}
where level 1 is the lower level, level 2 is  the upper level,
$n_2 $ is the gas number density, $n_2 A_{21}$ is the rate of emission
of photons from a unit volume, $A_{21}$ is the Einstein coefficient
for the transition, $h$ is the Planck constant and $\nu_{21}$ is
the frequency under consideration, see \cite{Hartigan2008}. In the case of
an optically thin medium, the intensity of the emission $I_{21} $ is
the integral along the line of sight
\begin{equation}
I_{21} = \int j_{21} dl
\quad .
\end {equation}
In the case of a constant gas number density
\begin{equation}
I_{21} \propto l
\quad ,
\end {equation}
where $l$ is the appropriate length, which
in astrophysical diffuse objects
depends
on the  orientation of the observer.

The numerical algorithm which allows us to build the image is now
outlined.
\begin{itemize}
\item An empty (value=0)
memory grid  ${\mathcal {M}} (i,j,k)$ which  contains
$NDIM^3$ pixels is considered
\item We  first  generate an
internal 3D surface by rotating the ideal image
 $180^{\circ}$
around the polar direction and a second  external  surface at a
fixed distance $\Delta R$ from the first surface. As an example,
we fixed $\Delta R$ = $ 0.03 R_{max}$, where $R_{max}$ is the
maximum radius of expansion, see \cite{Smith_2006}.
The points on
the memory grid which lie between the internal and external
surfaces are memorized on  ${\mathcal {M}} (i,j,k)$ (value=1).
\item Each point of
${\mathcal {M}} (i,j,k)$  has spatial coordinates $x,y,z$ which  can be
represented by the following $1 \times 3$  matrix, $A$,
\begin{equation}
A=
 \left[ \begin {array}{c} x \\\noalign{\medskip}y\\\noalign{\medskip}{
\it z}\end {array} \right]
\quad  .
\end{equation}
The orientation  of the object is characterized by
 the
Euler angles $(\Phi, \Theta, \Psi)$
and  therefore  by a total
 $3 \times 3$  rotation matrix,
$E$, see \cite{Goldstein2002}.
The matrix point  is
represented by the following $1 \times 3$  matrix, $B$,
\begin{equation}
B = E \cdot A
\quad .
\end{equation}
\item
The intensity map is obtained by summing the points of the
rotated images
along a particular direction.

\end{itemize}

\begin{figure}
  \begin{center}
\includegraphics[width=10cm]{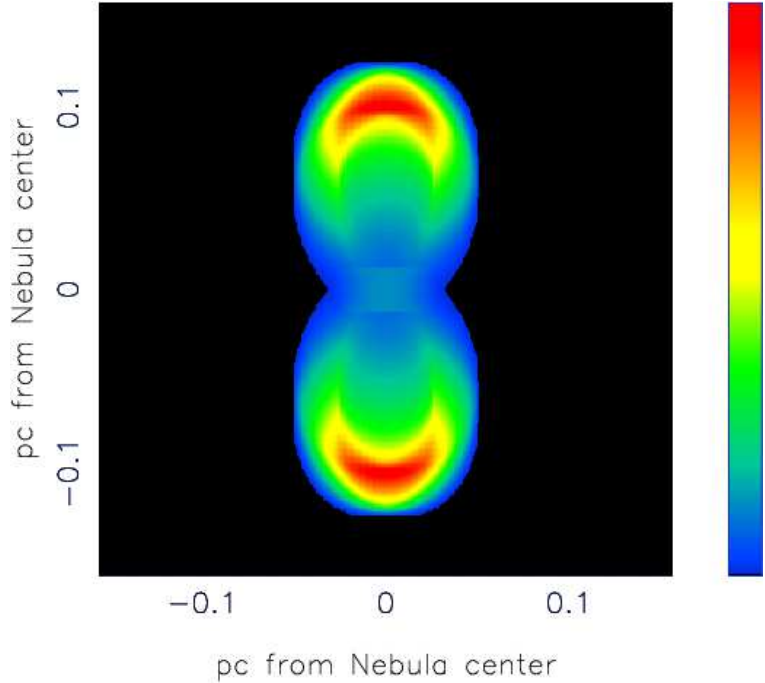}
  \end {center}
\caption {
Map of the theoretical intensity  of the Homunculus nebula
in the presence of an exponentially varying medium.
Physical parameters as in Table~\ref{parametershom}.
The three Euler angles
characterizing the   orientation
  are $ \Phi $=180$^{\circ }$,
$ \Theta $=90$^{\circ }$
and $ \Psi $=0$^{\circ }$.
This  combination of Euler angles corresponds
to the rotated image with the polar axis along the
z-axis.}%
    \label{eta_heat}
    \end{figure}

\begin{figure}
  \begin{center}
\includegraphics[width=10cm]{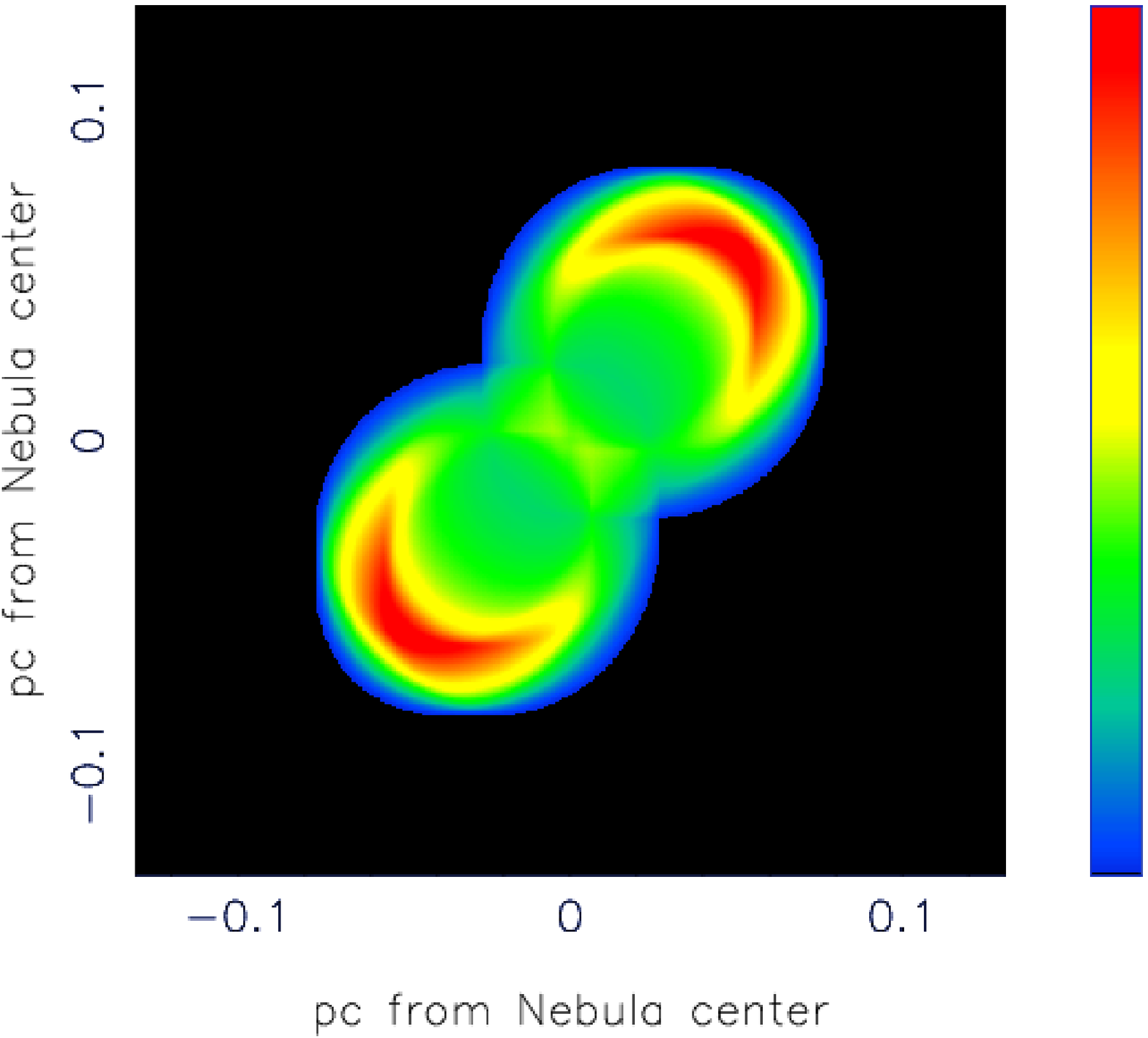}
  \end {center}
\caption {
Model map of the Homunculus nebula rotated in
accordance with the observations, for an exponentially
varying
medium.
Physical parameters as in Table~\ref{parametershom}.
The three Euler  angles characterizing
the orientation of the observer
are
     $ \Phi   $=130$^{\circ }$,
     $ \Theta $=40$^{\circ }$
and  $ \Psi   $=-140$^{\circ }$.
This  combination of Euler angles corresponds
to the observed image.
          }%
    \label{eta40_heat}
    \end{figure}

An  ideal image of the
Homunculus
nebula having the polar axis aligned with the z-direction
which means  polar axis along the z-direction,
is shown in Figure~\ref{eta_heat}
and this should be compared
with the $H_2$ emission structure reported
in Figure~4 of \cite{Smith2006}.
A model for a realistically rotated Homunculus
is shown in Figure \ref{eta40_heat}.
This should be compared with
Figure 1 in \cite{Smith2000} or
Figure 1 in \cite{Smith2006}.

\begin{figure*}
\begin{center}
\includegraphics[width=10cm]{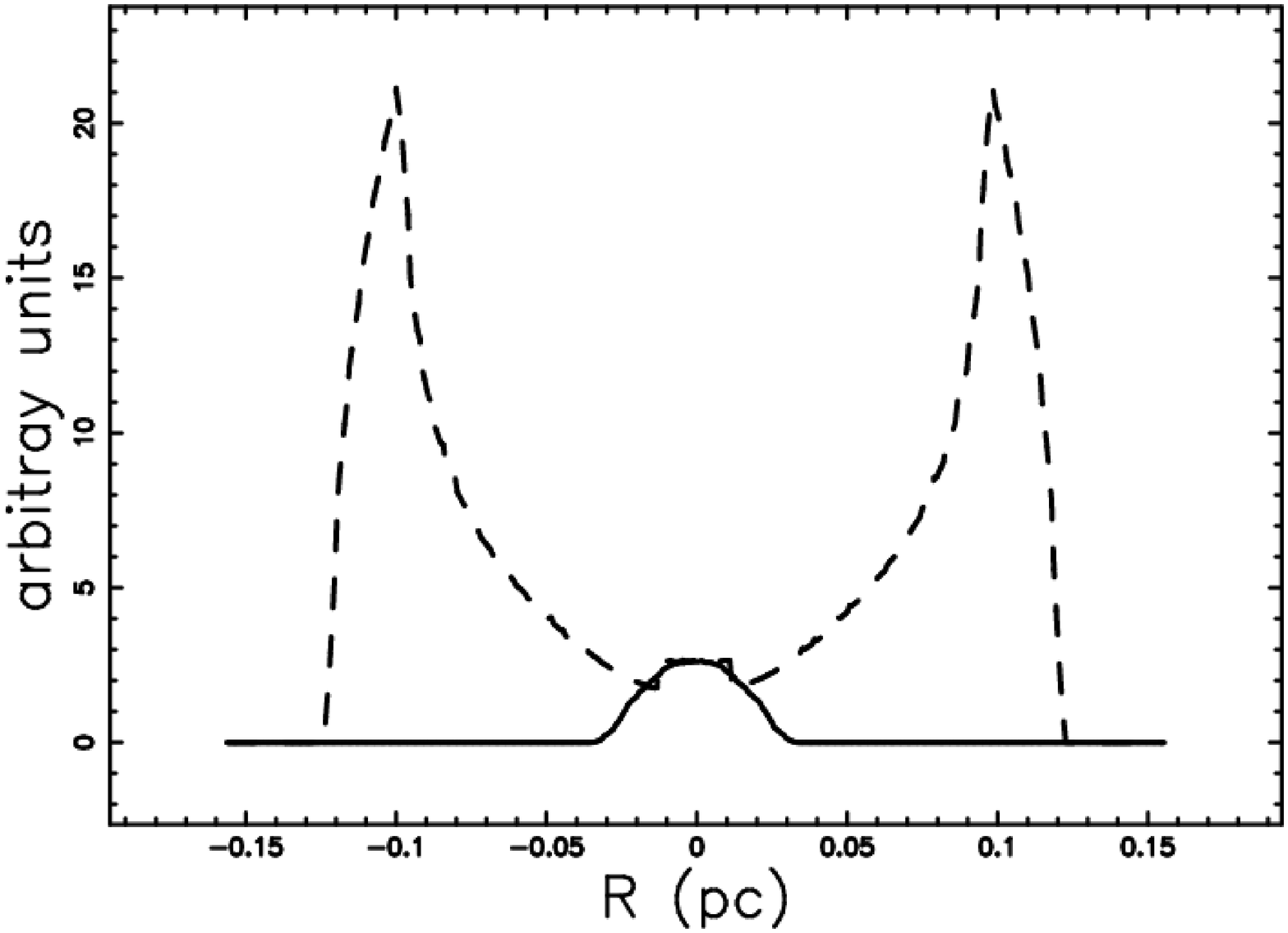}
\end {center}
\caption
{
 Two cuts of the model intensity
 across the center of the Homunculus nebula
 for  an exponentially varying medium:
 equatorial cut (full line)
 and polar cut  (dotted line).
 Parameters as in Figure~\ref{eta_heat}.
}
\label{cut_xy_eta}
    \end{figure*}

\begin{figure*}
\begin{center}
\includegraphics[width=10cm]{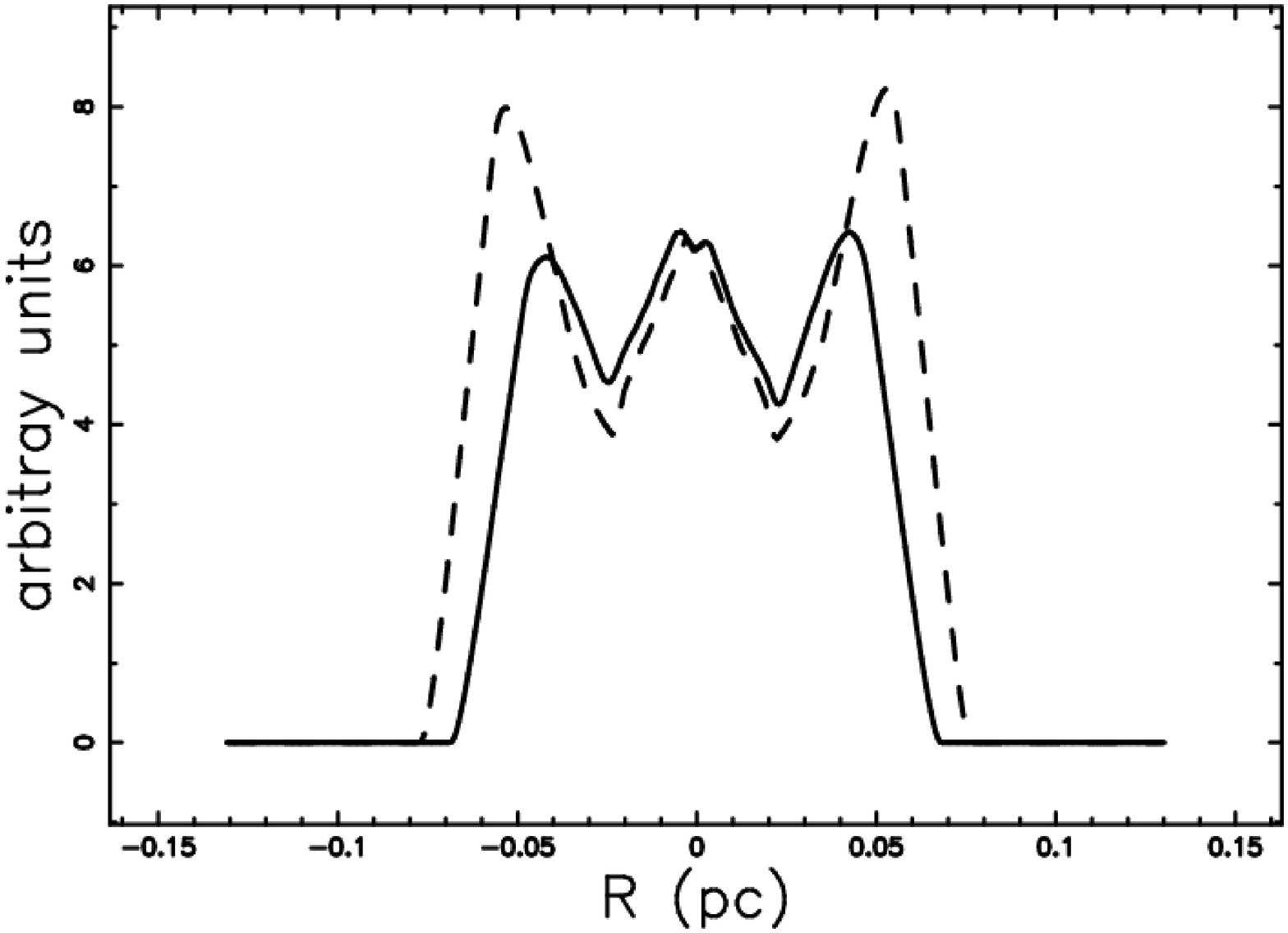}
\end {center}
\caption
{
 Two cuts of the model intensity
 across the center
 of the realistically rotated
 Homunculus nebula
 for  an exponentially varying medium:
 equatorial cut (full line)
 and polar cut  (dotted line).
 Parameters as in Figure~\ref{eta40_heat}.
}
\label{cut_xy_eta40}
    \end{figure*}

The rotated image exhibits
a double ring
and an intensity enhancement in the central
region which characterizes the little
Homunculus,
see~\cite{Smith2002,Ishibashi2003,Smith_2005,Gonzales_2006}.
Figure~\ref{cut_xy_eta} and Figure~\ref{cut_xy_eta40} show two
cuts through the Homunculus nebula
without and with rotation.
The intensity enhancement is due to a projection effect
and is an alternative for the theory that associates
the little Homunculus
with  an eruption occurring some time
after the Great Eruption, see
\cite{Ishibashi2003,Smith_2005}. We briefly recall that  a central
enhancement is visible in one of the various
morphologies characterizing planetary nebulae.
This can be compared with   the model $BL_1-F$ in Figure~3 of  the Atlas of
synthetic line profiles by \cite{Morisset2008}.

Such cuts are common when analyzing planetary nebulae.
 As an example  Figure 4 in \cite{Jacoby2001} reports a
nearly symmetrical profile of the intensity
in the [OIII] image of A39, a nearly
spherical planetary  nebula.
Another example is the east-west cut in
$H{\beta}$
for the elliptical Ring nebula, crossing
the center of the nebula, see Figure~1
in \cite{Garnett2001}.
Such intensity
cuts  are not yet available
for $\eta$-Carinae and therefore can represent a new target for the
observers.

\section{Conclusions}

{\bf Law of motion}
We have analyzed the law of
motion determined
by the conservation of radial momentum
under two circumstances:
 exponential
or power law variation
 of the density of the interstellar medium.
The two non-linear equations
of motion are given by formulae
(17) in \cite{Zaninetti2009a} and \ref{nl_power}).
A comparison of the observed and simulated profiles
(Tables \ref{efficiency_exp} and \ref{efficiency_power}),
shows that the profiles are better described by
an exponential variation than by a power law.
Here we have assumed that all the swept-up mass  resides in a
thin shell beyond the advancing surface.
In the case of clumpy material this
 approximation
can be refined by introducing the porosity p.
Then, the accumulated mass
is $M^{1/p}$ rather than $M$.

We briefly recall that the
general principles of mass addition to astrophysical
flows via hydrodynamic mixing
have  been investigated by  \cite{Hartquist1986} and
\cite{Pittard2007}.
In order to fix the porosity parameter $p$
the equation of motion along the polar direction
should be provided in the form
\begin{equation}
R(t) = r_{obs} t^{\alpha_{obs}}
\label{rpower}
\quad ,
\end{equation}
where the two parameters $r_{obs}$
and $\alpha_{obs}$ are
found
 from an analysis
 of the observational
data.
As an example  the data over a 10 year period
of SN\,1993J\,
can be approximated by  a power law dependence
of the type  $R\,  \propto  t^{0.82}$, see
\cite{Marcaide2009}.

{\bf Images}
Assuming an optically
thin medium,
it is
possible to make a model image
of the Homunculus nebula once two hypotheses are made:
\begin{enumerate}
\item
The thickness of the emitting layer, $\Delta R$,
is the same everywhere
$\Delta R = 0.03 R_{max} $, where $R_{max}$
 is the maximum radius of expansion.
\item
The density of the emitting layer is constant
everywhere
\end{enumerate}
A 2D image  of the Homunculus nebula is shown in
Figure~\ref{eta_heat}
and
a non-rotated image in Figure \ref{eta40_heat}.
Our model provides an
explanation for the emission maps in e.g. X-rays (0.2 -1.5 keV)
by CHANDRA.
An  inspection of Figure
1 in \cite{Corcoran2004}
shows
that
the next generation
of X-ray observatories may
resolve the double ring
as observed in the near infrared, see   Figure 4 of
\cite{Smith2006} and our simulation  in
Figure~\ref{eta40_heat}. The
validity of our assumptions
 can be
checked
by looking at cuts
 in intensity over the various bands.
We recall that cuts
in intensity allow to fix the thickness
of the swept-up layer of a planetary nebula such as
  $A39$
where$\frac{\Delta R} {R} \approx \frac {10^{\prime}}
{77^{\prime}}$, see \cite{Jacoby2001}.


\end{document}